\documentclass[twocolumn,showpacs,preprintnumbers,amsmath,amssymb,prb]{revtex4}

\usepackage{graphicx}
\usepackage{dcolumn}
\usepackage{bm}

\begin{document}

\preprint{}

\title{Optical spin pumping of modulation doped electrons \\ probed by a two-color Kerr rotation technique}

\author{H.~Hoffmann$^{1}$}
\author{G.~V.~Astakhov$^{1,2}$}\email[E-mail: ]{astakhov@physik.uni-wuerzburg.de}
\author{T.~Kiessling$^{1}$}
\author{W.~Ossau$^{1}$}
\author{G.~Karczewski$^{3}$}
\author{T.~Wojtowicz$^{3}$}
\author{J.~Kossut$^{3}$}
\author{L.~W.~Molenkamp$^{1}$}

\affiliation{ $^{1}$Physikalisches Institut (EP3), Universit\"{a}t
W\"{u}rzburg, 97074 W\"{u}rzburg, Germany \\
$^{2}$A.F.Ioffe Physico-Technical Institute, Russian Academy of
Sciences, 194021 St.Petersburg, Russia \\
$^{3}$Institute of Physics, Polish Academy of Sciences, 02668
Warsaw, Poland }

\date{\today}

\begin{abstract}
We report on optical spin pumping of modulation electrons in
CdTe-based quantum wells with low intrinsic electron density (by
$10^{10}$~cm$^{-2}$). Under continuous wave excitation, we reach a
steady state accumulated spin density of about $10^{8}$~cm$^{-2}$.
Using a two-color Hanle-MOKE technique, we find a spin relaxation
time of 34~ns for the localized electrons in the nearly
unperturbed electron gas. Independent variation of the pump and
probe energies demonstrates the presence of additional
non-localized electrons in the quantum well, whose spin relaxation
time is substantially shorter.
\end{abstract}

\pacs{72.25.Fe, 78.66.Hf, 76.30.Pk}

\maketitle

The idea to use the spin of electrons and nuclei rather than the
electron charge for information processing \cite{SpinTr} has
renewed the interest on spin-related phenomena in solids.
Spin-based concepts for semiconductor devices require the
preparation of a long-lived spin state.  Diluted magnetic
semiconductors (DMS) exhibit a giant Zeeman splitting, enabling
efficient spin selection and injection \cite{SpinInj1, SpinInj2,
RTD}. An advantage of the II-VI semiconductors is that the
magnetic impurities, such as Mn or Cr, are incorporated
\textit{isoelectronically}. This enables the fabrication of high
quality DMS structures whose magnetic and electronic properties
can be varied independently. II-VI semiconductors as ZnSe or CdTe,
whose growth procedure is well optimized, are thus well suited for
spin coherence studies. Most current work in this direction,
however, concentrates on GaAs. E.g., a spin memory of free
electrons in excess of 100~ns has been reported for bulk n-GaAs
\cite{Long-GaAs1, Long-GaAs2}. For II-VI's, the longest lived spin
polarization reported so far is by two orders of magnitude shorter
and was observed in heavily doped \cite{ZnSe} and later in undoped
\cite{ZnSe_add} ZnSe quantum wells (QWs).

In strongly conducting samples the D'yakonov-Perel' (DP)
mechanism\cite{OO} dominates the spin relaxation. This mechanism
is quenched for weakly doped, insulating samples when the
electrons are localized. In this case, the hyperfine interaction
with the nuclei may result in an extended spin
coherence,\cite{SpinMan} which can be controlled by
electron-electron interactions. Calculations show a non-monotonic
behavior for the spin relaxation time vs. doping concetration and
suggest that a maximum occurs in the intermediate doping regime,
at the onset of the insulator phase\cite{SpinMaxN}. QWs are
attractive for such studies as modulation doping provides
wide-range variation of the intrinsic electron concentration
$n_e$. The properties of the quasi-two-dimensional electron gas
(2DEG) can be monitored through the optical density of neutral
excitons ($\mathrm{X}$) and negatively charged trions
($\mathrm{T}$) \cite{Trion0, Trion1, Trion2}. For slightly doped
CdTe QWs, spin relaxation times of 19~ns
(Ref.~\onlinecite{Spin_CdTe1}) and 30~ns
(Ref.~\onlinecite{Spin_CdTe2}) in CdTe QWs have recently been
reported, exceeding those for comparable GaAs QWs \cite{GaAs_QW}.

In the present paper we report on further experiments on efficient spin
pumping of intrinsically present modulation doped electrons in
CdTe-based QWs. The samples used for our studies are insulating,
with a 2DEG density on the order of $10^{10}$~cm$^{-2}$. We use a
highly sensitive technique, basically a two-color magneto-optical
Kerr effect (MOKE) experiment in combination with a Hanle
experiment, to probe the net spin polarization under continuous wave
(cw) excitation\cite{Kerr1, Kerr2}. The Hanle effect(the decrease of
polarization under application of in-plane magnetic fields) gives a
characteristic magnetic field $B_{1/2}$ for the depolarization of
the 2DEG, which directly yields the spin relaxation time $\tau_s$
\cite{OO}. We find that the two-color Hanle-MOKE experiment is a
very sensitive tool for measuring spin relaxation and produces
an easily observable signal even at very weak illumination. The
technique avoids pulsed excitation and, thus, allows a determination
of $\tau_s$ for a nearly unperturbed 2DEG.

For cw pumping, a tunable dye-laser is used. The excitation is
modulated between $\sigma^{+}$ and $\sigma^{-}$ circular
polarizations at a frequency of 50~kHz using a photoelastic quartz
modulator. The degree of circular polarization of the
photoluminescence (PL) is detected by a Si-based avalanche
photodiode and a two channel photon counter. The net spin
polarization is probed using a linearly polarized cw Ti:sapphire
laser. The photoinduced Kerr rotation $\theta$ is measured by a
balanced diode detector and demodulated by a lock-in amplifier.
Both pump and probe beams are focused to the same $d \approx
300$~$\mathrm{\mu m}$-diameter spot. An external magnetic field
can be applied in the sample plane (Voigt geometry). All
experiments are carried out at a temperature of 1.8~K; the samples
are immersed in superfluid helium.

\begin{figure}[tbp]
\includegraphics[width=.37\textwidth]{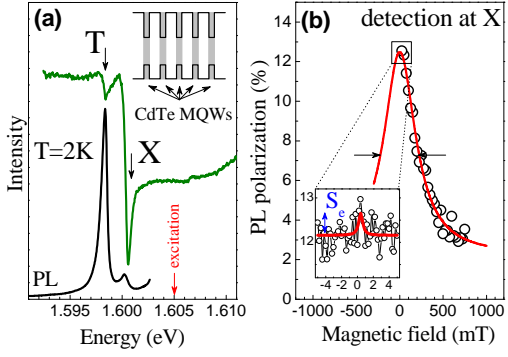}
\caption{(Color online) (a) Photoluminescence and reflectivity
spectra of 200-\AA-wide $\mathrm{CdTe/Cd_{0.78}Mg_{0.22}Te}$ MQWs.
PL is recorded under quasi-resonant excitation with energy
$E_{pump} = 1.605$~eV. The trion ($\mathrm{T}$) and exciton
($\mathrm{X}$) resonances are labeled by arrows. Inset: scheme of
the structure. (b) Hanle polarization data detected at the exciton
luminescence line. Solid line is a Lorentzian fit with
 $B_{1/2}' =220$~mT. Inset: enlarged scale for low fields.
$T=2$~K. $P_{pump}=20$~mW. $n_e = 1.3 \times 10^{10}$~cm$^{-2}$. }
\label{fig1}
\end{figure}

The samples have been grown by molecular-beam epitaxy on
(100)-oriented GaAs substrate. We present data for five 200-\AA-wide
$\mathrm{CdTe/Cd_{0.78}Mg_{0.22}Te}$ multiple quantum wells (MQWs)
 (see inset in Fig.\ref{fig1}a). The samples are
nominally undoped. The low density 2DEGs in the CdTe MQWs are due
to residual n-type doping of the $\mathrm{Cd_{0.78}Mg_{0.22}Te}$
barrier. Characteristic PL and reflectivity spectra in the
excitonic region of the MQWs are shown in Fig.~\ref{fig1}a. A pair
of resonances associated with the neutral exciton ($\mathrm{X}$)
and negatively charged trion ($\mathrm{T}$) is well resolved
\cite{Trion0,SingTrpl}. From the trion-to-exciton ratio of the
oscillator strength in the reflectivity spectrum we estimate the
concentration of modulation doped electrons at $n_e = 1.3 \times
10^{10}$~cm$^{-2}$ (Ref.~\onlinecite{Method}). The narrow
inhomogeneous broadening of 0.6~meV of the $\mathrm{X}$ and
$\mathrm{T}$ lines is indicative of the high quality of the
sample.

\begin{figure}[tbp]
\includegraphics[width=.37\textwidth]{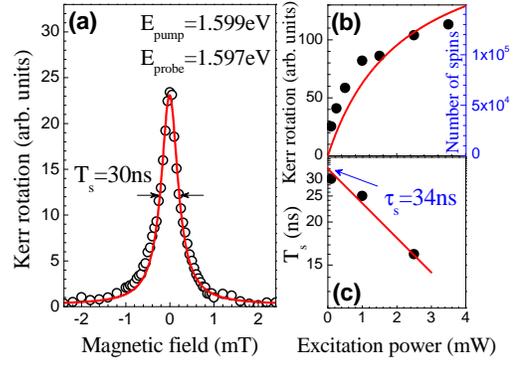}
\caption{(Color online) (a) The Hanle-MOKE signal, excited above
($E_{pump} = 1.599$~eV), and detected below the trion transition
($E_{probe} = 1.597$~eV) in the low power limit
($P_{pump}=0.1$~mW). The symbols are experimental points, the line
is a fit with Eq.~(\ref{HanleMOKE}). (b) The Kerr rotation angle
$\theta_i$ at zero applied field, which is proportional to the
number of spins $N_S$, as function of excitation power. The drawn
line is a fit using Eq.~(\ref{Pump}). (c) Spin life time $T_s$ as
function of excitation power. Following Eq.~(\ref{Ts}) one
extrapolates in the limit $P_{pump} \rightarrow 0$ to $\tau_s =
34$~ns. } \label{fig2}
\end{figure}

The key results of the optical spin pumping experiments are
collected in Fig.~\ref{fig2}. Panel (a) shows the Hanle-MOKE
signal for a pump energy slightly above, and a probe energy
slightly below the trion transition. The Kerr rotation angle
$\theta$ (and thus the spin polarization) completely vanishes with
magnetic field following a Lorentzian
\begin{equation}
\theta= \frac{\theta_i}{1+(B / B_{1/2})^2} \, \label{HanleMOKE}
\end{equation}
with $B_{1/2} = 0.23$~mT. Here, $\theta_i$ is proportional to the
number of pumped spins in zero magnetic field $B = 0$ and
increases with rising excitation power (Fig.~\ref{fig2}b).

Equation~(\ref{HanleMOKE}) allows us to determine $\tau_s$ from
the experiment. Application of an external magnetic field $B$ in
the sample plane results in spin precession of the electron spin
around the applied field with Larmor frequency $\omega_L = |g_e|
\mu_B B / \hbar$. Here, $g_e$ is electron g-factor. The time
evolution of the spin polarization upon delta pulse excitation can
be expressed as $\theta(t)=\theta_i \cos(\omega_L t) \exp (-t/
T_s)$, where the spin lifetime $T_s$ is related to, but not
necessarily equal to, the intrinsic spin memory time $\tau_s$. For
the cw limit one needs to integrate over time, which immediately
leads to Eq.~(\ref{HanleMOKE}) with $B / B_{1/2}= \omega_L T_s$
\cite{OO}. Thus, one directly obtains $T_s = \hbar / (|g_e| \mu_B
B_{1/2})$. With $g_e = - 1.64$,\cite{gFActor} the characteristic
magnetic field $B_{1/2} = 0.23$~mT corresponds to $T_s = 30$~ns
(Fig.~\ref{fig2}a). The exact relationship between the $T_s$ and
the spin relaxation time $\tau_s$ of the unperturbed 2DEG depends
on the pump power $P_{pump}$ and is discussed below. However, it
is obvious that $T_s \rightarrow \tau_s$ as $P_{pump} \rightarrow
0$. The data in Fig.~\ref{fig2}c suggest that $T_s$ obtained with
$P_{pump}=0.1$~mW corresponds to the low power limit with $\tau_s
\gtrsim 30$~ns.

Let us now describe the details of the spin pumping process.
Because of the optical selection rules, a circularly polarized
photon creates a spin polarized electron \cite{OO}. When these
polarized electrons replace the previously present unpolarized
electrons in the MQW, an effective spin pumping of the 2DEG
occurs. Under resonant excitation possible mechanisms are, e.g.,
spin-dependent formation of the trion singlet state, and electron
exchange scattering of the exciton state. A detailed consideration
of these mechanisms is beyond the scope of the present work.
Lumped together, these processes can be modeled by the
introduction of $S_e$, the maximum obtainable photoinduced spin
polarization of the 2DEG in the saturation regime. Following the
approach for n-GaAs (outlined in Refs.~\onlinecite{OO,
Long-GaAs2}), the spin pumping rate equations (at zero magnetic
field) can be written as
\begin{equation}
\frac{\partial (n_e S)}{\partial t} = G S_e - \frac{n_e}{\tau_J} S
- \frac{S}{\tau_s} n_e \,,\,\,\,\,\,\,\, \frac{\partial
n_e}{\partial t} = G - \frac{n_e}{\tau_J} \,. \label{Rate}
\end{equation}
The generation rate $G$ is proportional to the excitation power ($G
= \gamma P_{pump}$), where $\gamma$ is a coefficient that depends on
spot size and absorption efficiency. $S$ is the actual spin
polarization of the 2DEG at a given $G$. The time $\tau_J$
characterizes the spin transfer rate from the 2DEG back to the
exciton or trion reservoir. Under steady-state conditions
Eqs.~(\ref{Rate}) yield $\tau_J^{-1} = \gamma P_{pump} / n_e$ and
$S=S_e T_s / \tau_J$, where
\begin{equation}
T_s^{-1} = \tau_s^{-1} + \frac{\gamma P_{pump}}{n_e} \,.\label{Ts}
\end{equation}
Following Eq.~(\ref{Ts}), an extrapolation of $T_s$ to zero
$P_{pump}$ (solid line in Fig.~\ref{fig2}c) yields the intrinsic
spin relaxation time of the 2DEG, $\tau_s = 34$~ns.

\begin{figure}[tbp]
\includegraphics[width=.43\textwidth]{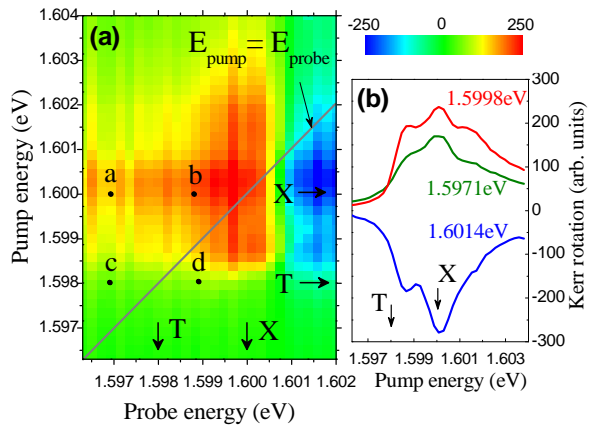}
\caption{(Color online) (a) 3D plot of the Kerr rotation angle
$\theta_i$ (at zero magnetic field) as function of pump
($E_{pump}$) and probe ($E_{probe}$) energies. The trion and
exciton transitions are indicated by arrows. Points a,b,c, and d
correspond to the similarly labeled panels in Fig.~\ref{fig4}. (b)
Spectra of the 2DEG spin polarization excitation (SPE) detected at
different energies ($ E_{probe}=1.5971, 1.5998, 1.6014$~eV).}
\label{fig3}
\end{figure}

The (zero field) Kerr rotation angle $\theta_i$ is proportional to
$\theta_i = \alpha S n_e$, where $\alpha$ is a function of the
detection energy $E_{probe}$. Collecting terms, we have
\begin{equation}
\theta_i = \alpha \frac{\gamma P_{pump} \tau_s}{n_e + \gamma
P_{pump} \tau_s} S_e n_e \,.\label{Pump}
\end{equation}
Eq.~(\ref{Pump}) describes the experimental dependence of $\theta_i$
on $P_{pump}$ very well. Fitting the experimental data in
Figs.~\ref{fig2}b and c, we find that $P_{pump} = 2.2 \, \mathrm{m
W}$ corresponds to $\gamma P_{pump} \tau_s = n_e$. In the arbitrary
units of Fig.~\ref{fig2}b, $\alpha S_e n_e$ corresponds to 200.

\begin{figure}[tbp]
\includegraphics[width=.41\textwidth]{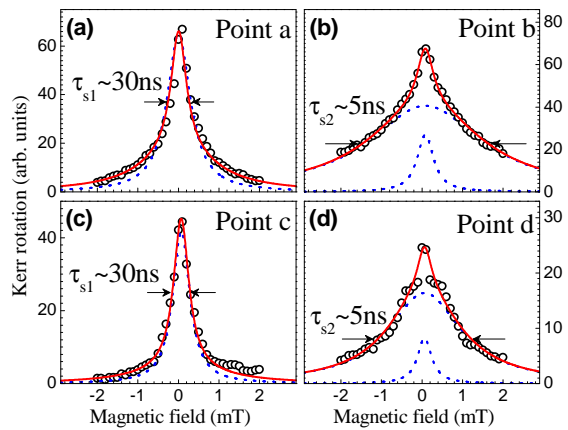}
\caption{(Color online) The Hanle-MOKE for different excitation
and detection energies (see also points in Fig.~\ref{fig3}a).
Symbols are experimental points, solid lines are fits with
Eq.~(\ref{HanleMOKE2}) and dotted lines are fits with
Eq.~(\ref{HanleMOKE}). Fitting parameters are collected in
Table~\ref{TabParam}. (a) $E_{pump} =1.600$~eV and $E_{probe} =
1.597$~eV. (b) $E_{pump} = 1.600$~eV and $E_{probe} = 1.599$~eV.
(c) $E_{pump} = 1.598$~eV and $E_{probe} = 1.597$~eV. (d)
$E_{pump} = 1.598$~eV and $E_{probe} = 1.599$~eV. Excitation and
detection power is $P_{pump}=0.25$~mW, $P_{probe}=0.17$~mW. }
\label{fig4}
\end{figure}

For comparison with our two-color technique, we also measured the
'classical' Hanle curve, analyzing the polarization of the PL at
the $\mathrm{X}$ emission line. The PL was excited 5~meV above the
exciton transition at $E_{pump} = 1.605$~eV (Fig.~\ref{fig1}a). As
the exciton binding energy exceeds 10~meV, excitons rather than
unbound electron-hole pairs are created. The Hanle data is shown
in Fig.~\ref{fig1}b and again is well described by
Eq.~(\ref{HanleMOKE}) but with $B_{1/2}' = 220$~mT. Note that in
this experiment the electron Hanle signal rides on a constant
background probably due to the field-independent hole
polarization. Formal using $T_s' = \hbar / (|g_e| \mu_B B_{1/2}')$
with $g_e = - 1.64$ yields $T_s' = 33$~ps. The difference between
$B_{1/2}$ and $B_{1/2}'$ by three orders of magnitude is not
surprising. An electron spin precession and relaxation in the
exciton may be strongly affected by interaction with a hole, and
thus it cannot be directly compared with that of an electron in
the 2DEG.

Nevertheless, the photoinduced spin polarization of the 2DEG,
$S_e$, also manifests itself in the 'classical' Hanle curve owing
to the spin dependent formation of the trion from the exciton.
This contribution reveals as a narrow peak appearing on a top of
the exciton Hanle curve, and has previously been observed in GaAs
QWs \cite{GaAs_QW}. However, upon testing many CdTe-based samples
\cite{Spin_CdTe1} we find that this peak is frequently weak. The
signal related to the 2DEG polarization in particular studied
samples is enlarged in the inset of Fig.~\ref{fig1}b. This data is
difficult to analyze and a value $S_e \sim 0.5$\% can only be
estimated as an upper limit. We use this value (obtained at
$P_{pump} = 20 \, \mathrm{mW}$) for deducing the effective number
of probed spins $N_S$, which is $N_S = 5 \pi (d/2)^2 S_e n_e$ at
high excitation power ($P_{pump} \gg 2.2 \, \mathrm{m W}$).
Assuming a spot size $d \sim 300$~$\mathrm{\mu m}$ we obtain $N_S
\sim 2 \times 10^5$. This value can be related to the saturation
level and thus enables calibration of our Kerr signal to a number
of spins (right axis in Fig.~\ref{fig2}b). Remarkably, the MOKE
significantly improves the sensitivity, allowing us to detect as
few as $10^{4}$ spins.

Rough estimate of the exciton recombination time $\tau_0 \sim
100$~ps \cite{commentTimes} allows us to obtain the average number
of excitons $\Delta n_X$ during our cw experiment, using $\Delta
n_X = G \tau_0$. At the characteristic pump power ($P_{pump} = 2.2
\, \mathrm{m W}$, Fig.~\ref{fig2}b) where $G \tau_s = n_e$, one
has $\Delta n_X = n_e \tau_0 / \tau_s$. The condition $\Delta n_X
\ll n_e$ is obviously fulfilled, which implies that the cw optical
pumping induces spin accumulation.

In an additional set of experiments, we measured the photoinduced
spin polarization of the 2DEG for different pump and probe
energies near the $\mathrm{X}$- and $\mathrm{T}$-transitions
(Fig.~\ref{fig3}a). As expected intuitively, the excitation
spectrum of the spin polarization (SPE), i.e. $\theta (E_{pump})$,
follows the optical density (reflectivity spectrum). This is shown
more clearly in the cross-section trace in Fig.~\ref{fig3}b. At
the same time, the Kerr rotation of the probe, $\theta
(E_{probe})$, has opposite sign on the high- and low energy sides.
This behavior is typical for the Kerr rotation when the probe
energy passes through a resonance \cite{Kerr_shape}.

In order to demonstrate the advantages of using independent
$E_{pump}$ and $E_{probe}$ energies (two-color mode), we measured
the Hanle-MOKE signal at the points labeled as a, b, c, and d in
Fig.~\ref{fig3}a. The resulting Hanle-MOKE curves are shown in
Fig.~\ref{fig4}. We find that in general there can be two
contributions to these curves and because of that some of them
cannot be described using Eq.~(\ref{HanleMOKE}) (dotted lines in
the figure). In order to explain this we assume (with
Ref.~\onlinecite{Ins-Con}) a spatially inhomogeneous distribution
of the electrons. The charged trions are more sensitive to
localization (for instance, in the electrostatic potential of
ionized donors in the barrier) as compared to their neutral
counterparts. As a result, when the detection energy is above the
$\mathrm{T}$-transition, only weakly localized electrons are
probed. For these electrons an alternative mechanism may dominate
the spin relaxation, resulting in a shortening of $\tau_s$.

\begin{table}
\caption{\label{TabParam} Parameters [in brackets] of fits to
Eq.~(\ref{HanleMOKE2})[(\ref{HanleMOKE})] shown as solid [dotted]
lines in Fig.~\ref{fig4}. Points a, b, c, and d refer to the
($E_{pump}$, $E_{probe}$) pairs indicated in Fig.~\ref{fig3}a. }
\begin{ruledtabular}
\begin{tabular}{ccccc}
& Point a & Point b & Point c & Point d \\
$E_{pump} =$  & 1.600~eV & 1.600~eV & 1.598~eV & 1.598~eV \\
$E_{probe} =$ & 1.597~eV & 1.599~eV & 1.597~eV & 1.599~eV \\
\hline
$\theta_{i}^{(2)} / \theta_{i}^{(1)}$ & 0.33 [0]    & 1.5 [-] & 0.08 [0]   & 1.98 [-] \\
$B_{1/2}^{(1)}$ (mT)          & 0.23 [0.35] & 0.24    & 0.23 [0.25]& 0.21     \\
$\tau_{s}^{(1)}$ (ns)      & 30 [20]     & 29       & 30 [28]   & 32        \\
$B_{1/2}^{(2)}$ (mT)          & 1.1         & 1.7     & 1.3       & 1.1      \\
$\tau_{s}^{(2)}$ (ns)      & 6           & 4       & 5          & 6       \\
\end{tabular}
\end{ruledtabular}
\end{table}

In order to fit all experimental data in Fig.~\ref{fig4} we extend
Eq.~(\ref{HanleMOKE}) to include two contributions:
\begin{equation}
\theta= \frac{\theta_{i}^{(1)}}{1+(B / B_{1/2}^{(1)})^2}  +
\frac{\theta_{i}^{(2)}}{1+(B /
B_{1/2}^{(2)})^2}\,,\label{HanleMOKE2}
\end{equation}
which allows us to fit all four cases well (solid lines in
Fig.~\ref{fig4}). The fitting parameters are given in
Table~\ref{TabParam}. We observe the following tendency in the
data: For detection below the trion transition (1.598~eV) at
(points a and c, $E_{probe} = 1.597$~eV) the spin relaxation time
is $\tau_{s}^{(1)} = 30 \pm 2$~ns. Shifting  the detection energy
only 2~meV higher (points b and d, $E_{probe} = 1.599$~eV) but
above the trion transition, we observe a dominant contribution
with a shorter spin relaxation time $\tau_{s}^{(2)} = 5 \pm 1$~ns,
which we tentatively assign to enhanced DP relaxation of
non-localized electrons. The relative weight of these
contributions (i.e., ratio $\theta_{i}^{(2)} / \theta_{i}^{(1)}$)
depends only slightly on the pump energy when the probe energy is
fixed.

In summary, we report on efficient spin pumping of modulation
doped electrons in CdTe QWs containing low density
($10^{10}$~cm$^{-2}$) insulating 2DEGs. It is monitored using a
technique based on the Kerr rotation, which is sensitive to
$10^{4}$ spins. We obtain a spin memory time $\tau_s = 34$~ns for
an unperturbed 2DEG. We find that our sample exhibits two electron
subsystems, whose spin relaxation time differs a factor six. This
observation confirms the importance of electron localization for
achieving long-lived spin coherence.

The authors thank R.~I.~Dzhioev, K.~V.~Kavokin, M.~V.~Lazarev and
especially V.~L.~Korenev for fruitful discussions. We also
acknowledge D.~R.~Yakovlev and E.~A.~Zhukov for providing us with
their data on time resolved Kerr rotation before publication
\cite{Spin_CdTe2}. This work was supported by the Deutsche
Forschungsgemeinschaft (SFB 410 and 436 RUS 113/843).

\end{document}